\documentclass[showpacs,showkeys,amsmath,amssymb,preprint]{revtex4}
\usepackage{epsfig,dcolumn,bm}

\begin{document}

\title{Kaon-nucleon interaction in the extended chiral SU(3) quark model}

\author{F. Huang}
\affiliation{ CCAST (World Laboratory), P.O. Box 8730, Beijing
100080, China \\ Institute of High Energy Physics, P.O. Box
918-4, Beijing 100049, China\footnote{Mailing address.} \\
Graduate School of the Chinese Academy of Sciences, Beijing,
China}
\author{Z.Y. Zhang}
\affiliation{Institute of High Energy Physics, P.O. Box 918-4,
Beijing 100049, China}

\begin{abstract}
The chiral SU(3) quark model is extended to include the coupling
between the quark and vector chiral fields. The one-gluon exchange
(OGE) which dominantly governs the short-range quark-quark
interaction in the original chiral SU(3) quark model is now nearly
replaced by the vector-meson exchange. Using this model, the
isospin $I=0$ and $I=1$ kaon-nucleon $S$, $P$, $D$, $F$ wave phase
shifts are dynamically studied by solving the resonating group
method (RGM) equation. Similar to those given by the original
chiral SU(3) quark model, the calculated results for many partial
waves are consistent with the experiment, while there is no
improvement in this new approach for the $P_{13}$ and $D_{15}$
channels, of which the theoretical phase shifts are too much
repulsive and attractive respectively when the laboratory momentum
of the kaon meson is greater than 300 MeV.
\end{abstract}

\pacs{13.75.Jz, 12.39.-x, 21.45.+v}

\keywords{$KN$ phase shifts; Quark model; Chiral symmetry}

\maketitle

\section{Introduction}

The kaon-nucleon ($KN$) scattering process has aroused particular
interest in the past and many works have been devoted to this
issue \cite{rbu90,dha02,barnes94,nbl02,bsi97,sle02,sle03,hjw03}.
In Ref. \cite{rbu90}, the J\"{u}lich group presented a
meson-exchange model on hadronic degrees of freedom to study the
$KN$ phase shifts. Considering single boson exchanges ($\sigma$,
$\rho$, and $\omega$) together with contributions from
higher-order diagrams involving $N$, $\Delta$, $K$, and $K^*$
intermediate states, the authors can give a good description of
$KN$ interaction, but the exchange of a short-range ($\sim$ 0.2
fm) phenomenological repulsive scalar meson $\sigma_{rep}$ had to
be added in order to reproduce the $S$-wave phase shifts in the
isospin $I=0$ channel. The range of this repulsion is much smaller
than the nucleon size, which clearly shows that the quark
substructure of the kaon and nucleon cannot be neglected. Further
in Ref. \cite{dha02} the authors refined this model by replacing
the phenomenological $\sigma_{rep}$ by one-gluon-exchange (OGE),
and a satisfactory description of the $KN$ experimental data was
gotten. However, in this hybrid model the one-pion exchange is
supposed to be absent, which is true on the hadron level, but is
not the case in a genuine quark model study, because the quark
exchange effect in the single boson exchanges has to be
considered. In Ref. \cite{barnes94}, Barnes and Swanson used the
quark-Born-diagram (QBD) method to derive the $KN$ scattering
amplitudes, and obtained reasonable results for the $KN$ phase
shifts, but it is limited to $S$-wave. Subsequently, the Born
approximation was applied to investigate the $KN$ scattering more
extensively in Ref. \cite{nbl02}. Nevertheless, the magnitudes of
most calculated phase shifts are too small. In Ref. \cite{bsi97},
taking the $\pi$ and $\sigma$ boson exchanges as well as the OGE
and confining potential as the quark-quark interactions, the
authors calculated the $S$-wave $KN$ phase shifts in a constituent
quark model by using the resonating group method (RGM). The
results are too attraction for $I=0$ channel and too repulsion for
$I=1$ channel, and thus the authors concluded that a consistent
description of $S$-wave $KN$ phase shifts in both isospin $I=0$
and $I=1$ channels simultaneously is not possible. In Ref.
\cite{sle02}, Lemaire {\it et al.} studied the $KN$ phase shifts
up to the orbit angular momentum $L=4$ on the quark level by using
the RGM method. They only considered the OGE and confining
potential as the quark-quark interaction, and their results can
give a reasonably description of the $S$-wave phase shifts, but
the $P$ and higher partial waves are poorly described. The authors
further incorporated $\pi$ and $\sigma$ exchanges besides the OGE
and confining potential in the quark-quark interaction in Ref.
\cite{sle03}, but the agreement obtained with the experimental
data is quite poor, especially the signs of the $S_{01}$,
$P_{03}$, $P_{11}$, $D_{05}$, $D_{13}$, $D_{15}$, $F_{07}$, and
$F_{15}$ waves are opposite to the experiment values. Recently,
Wang {\it et al.} \cite{hjw03} gave a study on the $KN$ elastic
scattering in a quark potential model. Their results are
consistent with the experimental data, but in their model, a
factor of color octet component is added arbitrarily and the size
parameter of harmonic oscillator is chosen to be $b_u=0.255$ fm,
which is too small compared with the radius of nucleon.

In spite of great successes, the constituent quark model needs to
have a logical explanation, from the underlying theory of the
strong interaction [i.e., Quantum Chromodynamics (QCD)] of the
source of the constituent quark mass. Thus spontaneous vacuum
breaking has to be considered, and as a consequence the coupling
between the quark field and the Goldstone boson is introduced to
restore the chiral symmetry. In this sense, the chiral quark model
can be regarded as a quite reasonable and useful model to describe
the medium-range nonperturbative QCD effect. By generalizing the
SU(2) linear $\sigma$ model, a chiral SU(3) quark model is
developed to describe the system with strangeness \cite{zyz97}.
This model has been quite successful in reproducing the energies
of the baryon ground states, the binding energy of deuteron, the
nucleon-nucleon ($NN$) scattering phase shifts of different
partial waves, and the hyperon-nucleon ($YN$) cross sections by
performing the RGM calculations \cite{zyz97,lrd03}. Inspired by
these achievements, we try to extend this model to study the
baryon-meson interactions. In our previous works
\cite{fhuang04nk,fhuang04dk}, we dynamically studied the $S$-,
$P$-, $D$-, and $F$-wave $KN$ phase shifts by performing a RGM
calculation. Comparing with Ref. \cite{sle03}, we obtained correct
signs of the phase shifts of $S_{01}$, $P_{11}$, $P_{03}$,
$D_{13}$, $D_{05}$, $F_{15}$, and $F_{07}$ partial waves, and for
$P_{01}$, $D_{03}$, and $D_{15}$ channels we also got a
considerable improvement in the magnitude. At the same time, the
satisfactory results also show that the short-range $KN$
interaction dominantly originates from the quark and one-gluon
exchanges.

It is a consensus that constituent quark is the dominat effective
degree of freedom for low-energy hadron physics, but about what
other proper effective degrees of freedom may be there still has
been a debate
\cite{glozman96,glozman00,isgur021,isgur022,liu99,liu00}. Glozman
and Riska proposed that the Goldstone boson is the only other
proper effective degree of freedom. In Ref.
\cite{glozman96,glozman00}, they applied the quark-chiral field
coupling model to study the baryon structure, and replaced OGE by
vector-meson coupling. They pointed out the spin-flavor
interaction is important in explaining the energy of the Roper
resonance and got a comparatively good fit to the baryon spectrum.
However Isgur gave a critique of the boson exchange model and
insisted that the OGE governs the baryon structure
\cite{isgur021,isgur022}. In Refs. \cite{liu99,liu00}, Liu {\it et
al.} produced a valence lattice QCD result which supports the
Goldstone boson exchange picture, but Isgur pointed out that this
is unjustified \cite{isgur021,isgur022}. On the other hand, in the
study of $NN$ interactions on the quark level, the short-range
feature can be explained by OGE interaction and quark exchange
effect, while in the traditional one-boson exchange (OBE) model on
the baryon level it comes from vector-meson ($\rho$, $K^*$,
$\omega$, and $\phi$) exchange. Some authors also studied the
short-range interaction as stemming from the Goldstone boson
exchanges on the quark level \cite{stancu97,shimizu00,lrd03}, and
it has been shown that these interactions can substitute
traditional OGE mechanism. Anyhow, for low-energy hadron physics,
what other proper effective degrees of freedom besides constituent
quarks may be, whether OGE or vector-meson exchange is the right
mechanism for describing the short-range quark-quark interaction,
or both of them are important, is still a controversial and
challenging problem.

In this paper, we extend the chiral SU(3) quark model to include
the coupling between the quark and vector chiral fields. The OGE
which dominantly governs the short-range quark-quark interaction
in the original chiral SU(3) quark model is now nearly replaced by
the vector-meson exchange. As we did in Refs.
\cite{fhuang04nk,fhuang04dk}, the mass of the $\sigma$ meson is
taken to be $675$ MeV and the mixing of $\sigma_0$ (scalar
singlet) and $\sigma_8$ (scalar iso-scalar) is considered. The set
of parameters we used can satisfactorily reproduce the energies of
the ground states of the octet and decuplet baryons. Using this
model, we perform a dynamical calculation of the $S$, $P$, $D$,
$F$ wave $KN$ phase shifts in the isospin $I=0$ and $I=1$ channels
by solving a RGM equation. The calculated phase shifts for
different partial waves are similar to those obtained by the
original chiral SU(3) quark model. In comparison with a recent RGM
study on a quark level \cite{sle03}, our investigation achieves a
considerable improvement on the theoretical phase shifts, and for
many channels the theoretical results are qualitatively consistent
with the experimental data. Nevertheless there is no improvement
in this new approach for the $P_{13}$ and $D_{15}$ partial waves,
of which the calculated phase shifts are too much repulsive and
attractive respectively when the laboratory momentum of the kaon
meson is greater than 300 MeV, as it was the case in the past. It
would be studied in future work if there are some physical
ingredients missing in our quark model investigations.

The paper is organized as follows. In the next section the
framework of the extended chiral SU(3) quark model is briefly
introduced. The results for the $S$-, $P$-, $D$-, and $F$-wave
$KN$ phase shifts are shown in Sec. III, where some discussion is
presented as well. Finally, the summary is given in Sec. IV.

\section{Formulation}

\subsection{Model}

The chiral SU(3) quark model has been widely described in the
literature \cite{fhuang04nk,fhuang04dk} and we refer the reader to
those works for details. Here we just give the salient features of
the extended chiral SU(3) quark model.

In the extended chiral SU(3) quark model, besides the nonet
pseudoscalar meson fields and nonet scalar meson fields, the
couplings among vector meson fields with quarks is also
considered. With this generalization, in the interaction
Lagrangian a term of coupling between the quark and vector meson
field is introduced,
\begin{eqnarray}
{\cal L}_I^v = -g_{chv} \bar{\psi}\gamma_\mu T^a A^\mu_a \psi
-\frac{f_{chv}}{2M_P} \bar{\psi} \sigma_{\mu\nu} T^a \partial^\nu
A^\mu_a \psi.
\end{eqnarray}
Thus the meson fields induced effective quark-quark potentials can
be written as
\begin{eqnarray}
V^{ch}_{ij} = \sum_{a=0}^8 V_{\sigma_a}({\bm r}_{ij})+\sum_{a=0}^8
V_{\pi_a}({\bm r}_{ij})+\sum_{a=0}^8 V_{\rho_a}({\bm r}_{ij}),
\end{eqnarray}
where $\sigma_{0},...,\sigma_{8}$ are the scalar nonet fields,
$\pi_{0},..,\pi_{8}$ the pseudoscalar nonet fields, and
$\rho_{0},..,\rho_{8}$ the vector nonet fields. The expressions of
these potentials are
\begin{eqnarray}
V_{\sigma_a}({\bm r}_{ij})=-C(g_{ch},m_{\sigma_a},\Lambda)
X_1(m_{\sigma_a},\Lambda,r_{ij}) [\lambda_a(i)\lambda_a(j)] +
V_{\sigma_a}^{\bm {l \cdot s}}({\bm r}_{ij}),
\end{eqnarray}
\begin{eqnarray}
V_{\pi_a}({\bm r}_{ij})=C(g_{ch},m_{\pi_a},\Lambda)
\frac{m^2_{\pi_a}}{12m_{q_i}m_{q_j}} X_2(m_{\pi_a},\Lambda,r_{ij})
({\bm \sigma}_i\cdot{\bm \sigma}_j) [\lambda_a(i)\lambda_a(j)]
+V_{\pi_a}^{ten}({\bm r}_{ij}),
\end{eqnarray}
\begin{eqnarray}
V_{\rho_a}({\bm r}_{ij})&=&C(g_{chv},m_{\rho_a},\Lambda)\left\{
X_1(m_{\rho_a},\Lambda,r_{ij})+
\frac{m^2_{\rho_a}}{6m_{q_i}m_{q_j}}
\left(1+\frac{f_{chv}}{g_{chv}}\frac{m_{q_i}+m_{q_j}}{M_P}+\frac{f^2_{chv}}{g^2_{chv}}
\right. \right. \nonumber \\
&& \left. \left. \times \frac{m_{q_i}m_{q_j}}{M^2_P}\right)
X_2(m_{\rho_a},\Lambda,r_{ij})({\bm \sigma}_i\cdot{\bm \sigma}_j)
\right\}[\lambda_a(i)\lambda_a(j)] + V_{\rho_a}^{\bm {l \cdot
s}}({\bm r}_{ij}) + V_{\rho_a}^{ten}({\bm r}_{ij}),
\end{eqnarray}
with
\begin{eqnarray}
V_{\sigma_a}^{\bm {l \cdot s}}({\bm r}_{ij})&=&
-C(g_{ch},m_{\sigma_a},\Lambda)\frac{m^2_{\sigma_a}}{4m_{q_i}m_{q_j}}
\left\{G(m_{\sigma_a}r_{ij})-\left(\frac{\Lambda}{m_{\sigma_a}}\right)^3
G(\Lambda r_{ij})\right\} \nonumber\\
&&\times[{\bm L \cdot ({\bm \sigma}_i+{\bm
\sigma}_j)}][\lambda_a(i)\lambda_a(j)],
\end{eqnarray}
\begin{eqnarray}
V_{\rho_a}^{\bm {l \cdot s}}({\bm r}_{ij})&=&
-C(g_{chv},m_{\rho_a},\Lambda)\frac{3m^2_{\rho_a}}{4m_{q_i}m_{q_j}}
\left(1+\frac{f_{chv}}{g_{chv}}\frac{2(m_{q_i}+m_{q_j})}{3M_P}\right)
\nonumber \\
&&\times
\left\{G(m_{\rho_a}r_{ij})-\left(\frac{\Lambda}{m_{\rho_a}}\right)^3
G(\Lambda r_{ij})\right\}[{\bm L \cdot ({\bm \sigma}_i+{\bm
\sigma}_j)}][\lambda_a(i)\lambda_a(j)],
\end{eqnarray}
and
\begin{eqnarray}
V_{\pi_a}^{ten}({\bm r}_{ij})=
C(g_{ch},m_{\pi_a},\Lambda)\frac{m^2_{\pi_a}}{12m_{q_i}m_{q_j}}
\left\{H(m_{\pi_a}r_{ij})-\left(\frac{\Lambda}{m_{\pi_a}}\right)^3
H(\Lambda r_{ij})\right\}\hat{S}_{ij}[\lambda_a(i)\lambda_a(j)],
\end{eqnarray}
\begin{eqnarray}
V_{\rho_a}^{ten}({\bm r}_{ij}) &=& -C(g_{chv},m_{\rho_a},\Lambda)
\frac{m^2_{\rho_a}}{12m_{q_i}m_{q_j}}
\left(1+\frac{f_{chv}}{g_{chv}}\frac{m_{q_i}+m_{q_j}}{M_P}+\frac{f^2_{chv}}{g^2_{chv}}\frac{m_{q_i}m_{q_j}}{M^2_P}\right)
\nonumber \\
&&\times\left\{H(m_{\pi_a}r_{ij})-\left(\frac{\Lambda}{m_{\pi_a}}\right)^3
H(\Lambda r_{ij})\right\}\hat{S}_{ij}[\lambda_a(i)\lambda_a(j)],
\end{eqnarray}
where
\begin{eqnarray}
C(g_{ch},m,\Lambda)=\frac{g^2_{ch}}{4\pi}
\frac{\Lambda^2}{\Lambda^2-m^2} m,
\end{eqnarray}
\begin{eqnarray}
\label{x1mlr} X_1(m,\Lambda,r)=Y(mr)-\frac{\Lambda}{m} Y(\Lambda
r),
\end{eqnarray}
\begin{eqnarray}
X_2(m,\Lambda,r)=Y(mr)-\left(\frac{\Lambda}{m}\right)^3 Y(\Lambda
r),
\end{eqnarray}
\begin{eqnarray}
Y(x)=\frac{1}{x}e^{-x},
\end{eqnarray}
\begin{eqnarray}
G(x)=\frac{1}{x}\left(1+\frac{1}{x}\right)Y(x),
\end{eqnarray}
\begin{eqnarray}
H(x)=\left(1+\frac{3}{x}+\frac{3}{x^2}\right)Y(x),
\end{eqnarray}
\begin{eqnarray}
\hat{S}_{ij}=\left[3({\bm \sigma}_i \cdot \hat{r}_{ij})({\bm
\sigma}_j \cdot \hat{r}_{ij})-{\bm \sigma}_i \cdot {\bm
\sigma}_j\right],
\end{eqnarray}
and $M_P$ being a mass scale, taken as proton mass. $m_{\sigma_a}$
is the mass of the scalar meson, $m_{\pi_a}$ the mass of the
pseudoscalar meson, and $m_{\rho_a}$ the mass of the vector meson.

For the systems with an antiquark ${\bar s}$, the total
Hamiltonian can be written as
\begin{eqnarray}
\label{hami5q}
H=\sum_{i=1}^{5}T_{i}-T_{G}+\sum_{i<j=1}^{4}V_{ij}+\sum_{i=1}^{4}V_{i\bar
5},
\end{eqnarray}
where $T_G$ is the kinetic energy operator for the center-of-mass
motion, and $V_{ij}$ and $V_{i\bar 5}$ represent the quark-quark
($qq$) and quark-antiquark ($q{\bar q}$) interactions,
respectively,
\begin{eqnarray}
V_{ij}= V^{OGE}_{ij} + V^{conf}_{ij} + V^{ch}_{ij},
\end{eqnarray}
where $V^{OGE}_{ij}$ is the one-gluon-exchange interaction,
\begin{eqnarray}
V^{OGE}_{ij}=\frac{1}{4}g_{i}g_{j}\left(\lambda^c_i\cdot\lambda^c_j\right)
\left\{\frac{1}{r_{ij}}-\frac{\pi}{2} \delta({\bm r}_{ij})
\left(\frac{1}{m^2_{q_i}}+\frac{1}{m^2_{q_j}}+\frac{4}{3}\frac{1}{m_{q_i}m_{q_j}}
({\bm \sigma}_i \cdot {\bm \sigma}_j)\right)\right\}+V_{OGE}^{\bm
l \cdot \bm s},
\end{eqnarray}
with
\begin{eqnarray}
V_{OGE}^{\bm l \cdot \bm
s}=-\frac{1}{16}g_ig_j\left(\lambda^c_i\cdot\lambda^c_j\right)
\frac{3}{m_{q_i}m_{q_j}}\frac{1}{r^3_{ij}}{\bm L \cdot ({\bm
\sigma}_i+{\bm \sigma}_j)},
\end{eqnarray}
and $V^{conf}_{ij}$ is the confinement potential, taken as the
quadratic form,
\begin{eqnarray}
V_{ij}^{conf}=-a_{ij}^{c}(\lambda_{i}^{c}\cdot\lambda_{j}^{c})r_{ij}^2
-a_{ij}^{c0}(\lambda_{i}^{c}\cdot\lambda_{j}^{c}).
\end{eqnarray}
$V_{i \bar 5}$ in Eq. (\ref{hami5q}) includes two parts: direct
interaction and annihilation parts,
\begin{eqnarray}
V_{i\bar 5}=V^{dir}_{i\bar 5}+V^{ann}_{i\bar 5},
\end{eqnarray}
with
\begin{eqnarray}
V_{i\bar 5}^{dir}=V_{i\bar 5}^{conf}+V_{i\bar 5}^{OGE}+V_{i\bar
5}^{ch},
\end{eqnarray}
where
\begin{eqnarray}
V_{i\bar
5}^{conf}=-a_{i5}^{c}\left(-\lambda_{i}^{c}\cdot{\lambda_{5}^{c}}^*\right)r_{i\bar
5}^2
-a_{i5}^{c0}\left(-\lambda_{i}^{c}\cdot{\lambda_{5}^{c}}^*\right),
\end{eqnarray}
\begin{eqnarray}
V^{OGE}_{i\bar
5}&=&\frac{1}{4}g_{i}g_{s}\left(-\lambda^c_i\cdot{\lambda^c_5}^*\right)
\left\{\frac{1}{r_{i\bar 5}}-\frac{\pi}{2} \delta({\bm r}_{i\bar
5})
\left(\frac{1}{m^2_{q_i}}+\frac{1}{m^2_{s}}+\frac{4}{3}\frac{1}{m_{q_i}m_{s}}
({\bm \sigma}_i \cdot {\bm \sigma}_5)\right)\right\}  \nonumber \\
&&-\frac{1}{16}g_ig_s\left(-\lambda^c_i\cdot{\lambda^c_5}^*\right)
\frac{3}{m_{q_i}m_{q_5}}\frac{1}{r^3_{i\bar 5}}{\bm L \cdot ({\bm
\sigma}_i+{\bm \sigma}_5)},
\end{eqnarray}
and
\begin{eqnarray}
V_{i\bar{5}}^{ch}=\sum_{j}(-1)^{G_j}V_{i5}^{ch,j}.
\end{eqnarray}
Here $(-1)^{G_j}$ represents the G parity of the $j$th meson. For
the $NK$ system, $u(d)\bar{s}$ can only annihilate into $K$ and
$K^*$ mesons---i.e.,
\begin{eqnarray}
V_{i\bar 5}^{ann}=V_{ann}^{K}+V_{ann}^{K^*},
\end{eqnarray}
with
\begin{eqnarray}
V_{ann}^{K}=C^K\left(\frac{1-{\bm \sigma}_q \cdot {\bm
\sigma}_{\bar{q}}}{2}\right)_{s}\left(\frac{2 + 3\lambda_q \cdot
\lambda^*_{\bar{q}}}{6}\right)_{c} \left(\frac{38+3\lambda_q \cdot
\lambda^*_{\bar q}}{18}\right)_{f}\delta({\bm r}),
\end{eqnarray}
and
\begin{eqnarray}
V_{ann}^{K^*}=C^{K^*}\left(\frac{3+{\bm \sigma}_q \cdot {\bm
\sigma}_{\bar{q}}}{2}\right)_{s}\left(\frac{2 + 3\lambda_q \cdot
\lambda^*_{\bar{q}}}{6}\right)_{c} \left(\frac{38+3\lambda_q \cdot
\lambda^*_{\bar q}}{18}\right)_{f}\delta({\bm r}),
\end{eqnarray}
where $C^K$ and $C^{K^*}$ are treated as parameters and we adjust
them to fit the mass of $K$ and $K^*$ mesons.

\subsection{Determination of the parameters}

{\small
\begin{table}[htb]
\caption{\label{para} Model parameters. The meson masses and the
cutoff masses: $m_{a_0}=980$ MeV, $m_{\kappa}=1430$ MeV,
$m_{f_0}=980$ MeV, $m_{\pi}=138$ MeV, $m_K=495$ MeV,
$m_{\eta}=549$ MeV, $m_{\eta'}=957$ MeV, $m_{\rho}=770$ MeV,
$m_{K^*}=892$ MeV, $m_{\omega}=782$ MeV, $m_{\phi}=1020$ MeV,
$\Lambda=1500$ MeV for $\kappa$ and 1100 MeV for other mesons.}
\begin{center}
\begin{tabular*}{160mm}{@{\extracolsep\fill}ccccc}
\hline\hline
  & \multicolumn{2}{c}{Chiral SU(3) quark model} & \multicolumn{2}{c}{Extended chiral SU(3) quark model} \\ \cline{2-3} \cline{4-5}
  & I & II & I & II \\
  & $\theta^S=35.264^\circ$ & $\theta^S=-18^\circ$ & $\theta^S=35.264^\circ$ & $\theta^S=-18^\circ$ \\
\hline
 $b_u$ (fm)  & 0.5 & 0.5 & 0.45 & 0.45 \\
 $m_u$ (MeV) & 313 & 313 & 313 & 313 \\
 $m_s$ (MeV) & 470 & 470 & 470 & 470 \\
 $g_u^2$     & 0.7704 & 0.7704 & 0.0748 & 0.0748 \\
 $g_s^2$    & 0.5525 & 0.5525 & 0.0001 & 0.0001 \\
 $g_{ch}$    & 2.621 & 2.621 & 2.621 & 2.621 \\
 $g_{chv}$   &  0    &  0    & 2.351 & 2.351 \\
 $m_\sigma$ (MeV) & 675 & 675 & 675 & 675 \\
 $a^c_{uu}$ (MeV/fm$^2$) & 52.9 & 55.7 & 56.4 & 60.3 \\
 $a^c_{us}$ (MeV/fm$^2$) & 76.0 & 72.1 & 104.1 & 98.8 \\
 $a^{c0}_{uu}$ (MeV)  & $-$51.7 & $-$56.4 & $-$86.4 & $-$91.8 \\
 $a^{c0}_{us}$ (MeV)  & $-$68.5 & $-$63.0 & $-$123.1 & $-$116.8 \\
\hline\hline
\end{tabular*}
\end{center}
\end{table}}

The harmonic-oscillator width parameter $b_u$ is taken with
different values for the two models: $b_u=0.50$ fm in the chiral
SU(3) quark model and $b_u=0.45$ fm in the extended chiral SU(3)
quark model. This means that the bare radius of baryon becomes
smaller when more meson clouds are included in the model, which
sounds reasonable in the sense of the physical picture. The up
(down) quark mass $m_{u(d)}$ and the strange quark mass $m_s$ are
taken to be the usual values: $m_{u(d)}=313$ MeV and $m_s=470$
MeV. The coupling constant for scalar and pseudoscalar chiral
field coupling, $g_{ch}$, is determined according to the relation
\begin{eqnarray}
\frac{g^{2}_{ch}}{4\pi} = \left( \frac{3}{5} \right)^{2}
\frac{g^{2}_{NN\pi}}{4\pi} \frac{m^{2}_{u}}{M^{2}_{N}},
\end{eqnarray}
with empirical value $g^{2}_{NN\pi}/4\pi=13.67$. $g_{chv}$ and
$f_{chv}$ are the coupling constants for vector coupling and
tensor coupling of the vector meson field, respectively. In the
study of nucleon resonance transition coupling to vector meson,
Riska and Brown took $g_{chv}=3.0$ and neglected the tensor
coupling part \cite{riska01}. From the one-boson exchange theory
on the baryon level, we can also obtain these two values according
to the SU(3) relation between quark and baryon levels. For
example,
\begin{eqnarray}
g_{chv}=g_{NN\rho},
\end{eqnarray}
\begin{eqnarray}
f_{chv}=\frac{3}{5}(f_{NN\rho}-4g_{NN\rho}).
\end{eqnarray}
In the Nijmegen model D, $g_{NN\rho}=2.09$ and $f_{NN\rho}=17.12$
\cite{nagels75}. From the two equations above, we get
$g_{chv}=2.09$ and $f_{chv}=5.26$. In this work, we neglect the
tensor coupling part of the vector meson field as did by Riska and
Brown \cite{riska01}, and take the coupling constant for vector
coupling of the vector-meson field to be $g_{chv}=2.351$ as the
same we used in the $NN$ scattering calculation \cite{lrd03},
which is a little bit smaller than the value used in Ref.
\cite{riska01}, but slightly larger than the value obtained from
the $NN\rho$ coupling constant of Nijmegen model D
\cite{nagels75}. The masses of all the mesons are taken to be the
experimental values, except for the $\sigma$ meson, whose mass is
treated as an adjustable parameter. We chose $m_{\sigma}=675$ MeV
as the same in the original chiral SU(3) quark model
\cite{fhuang04nk}, where it is fixed by the $S$-wave $KN$ phase
shifts. The cutoff radius $\Lambda^{-1}$ is taken to be the value
close to the chiral symmetry breaking scale
\cite{ito90,amk91,abu91,emh91}. After the parameters of chiral
fields are fixed, the coupling constants of OGE, $g_{u}$ and
$g_{s}$, can be determined by the mass splits between $N$,
$\Delta$ and $\Sigma$, $\Lambda$, respectively. The confinement
strengths $a^{c}_{uu}$, $a^{c}_{us}$, and $a^{c}_{ss}$ are fixed
by the stability conditions of $N$, $\Lambda$, and $\Xi$ and the
zero-point energies $a^{c0}_{uu}$, $a^{c0}_{us}$, and
$a^{c0}_{ss}$ by fitting the masses of $N$, $\Sigma$, and
$\overline{\Xi+\Omega}$, respectively.

{\small
\begin{table}[htb]
\caption{\label{baryonmass} The masses of octet and decuplet
baryons.}
\begin{tabular*}{165mm}{@{\extracolsep\fill}lcccccccc}
\hline\hline
       & $N$ & $\Sigma$ & $\Xi$ & $\Lambda$ & $\Delta$ & $\Sigma^\ast$ & $\Xi^\ast$ & $\Omega$  \\
\hline
Theor. & 939 &   1194   & 1335  &   1116    &    1232  &       1370    &    1511    &   1684    \\
Expt.  & 939 &   1194   & 1319  &   1116    &    1232  &       1385    &    1530    &   1672    \\
\hline\hline
\end{tabular*}
\end{table}}

In the calculation, $\eta$ and $\eta'$ mesons are mixed by
$\eta_1$ and $\eta_8$ with the mixing angle $\theta^{PS}$ taken to
be the usual value $-23^\circ$. $\omega$ and $\phi$ mesons consist
of $\sqrt{1/2}(u{\bar u}+d{\bar d})$ and $(s{\bar s})$,
respectively, i.e., they are ideally mixed by $\omega_1$ and
$\omega_8$ with the mixing angle $\theta^V=35.264^\circ$. For the
$KN$ case, we also consider the mixing between $\sigma_0$ and
$\sigma_8$. The mixing angle $\theta^{S}$ is an open issue because
the structure of the $\sigma$ meson is still unclear and
controversial. We adopt two possible values as in our previous
works \cite{fhuang04nk,fhuang04dk}, one is $35.264^\circ$ which
means that $\sigma$ and $f_0$ [In our previous works $f_0$ was
named $\epsilon$ and $a_0$ was named $\sigma'$] are ideally mixed
by $\sigma_0$ and $\sigma_8$, and the other is $-18^\circ$ which
is provided by Dai and Wu based on their recent investigation of a
dynamically spontaneous symmetry breaking mechanism \cite{ybd03}.
In both of these two cases, the attraction of the $\sigma$ meson
can be reduced a lot, and thus we can get reasonable $S$-wave $KN$
phase shifts.

The model parameters are summarized in Table \ref{para}. The
masses of octet and decuplet baryons obtained from the extended
chiral SU(3) quark model are listed in Table \ref{baryonmass}.

\subsection{Dynamical study of the $KN$ phase shifts}

With all parameters determined in the extended chiral SU(3) quark
model, the $KN$ phase shifts can be dynamically studied in the
frame work of the RGM. The wave function of the five quark system
is of the following form:
\begin{eqnarray}
\Psi={\cal A}[{\hat \phi}_N(\bm \xi_1,\bm \xi_2) {\hat \phi}_K(\bm
\xi_3) \chi({\bm R}_{NK})],
\end{eqnarray}
where ${\bm \xi}_1$ and ${\bm \xi}_2$ are the internal coordinates
for the cluster $N$, and ${\bm \xi}_3$ the internal coordinate for
the cluster $K$. ${\bm R}_{NK}\equiv {\bm R}_N-{\bm R}_K$ is the
relative coordinate between the two clusters, $N$ and $K$. The
${\hat \phi}_N$ is the antisymmetrized internal cluster wave
function of $N$, and $\chi({\bm R}_{NK})$ the relative wave
function of the two clusters. The symbol $\cal A$ is the
antisymmetrizing operator defined as
\begin{equation}
{\cal A}\equiv{1-\sum_{i \in N}P_{i4}}\equiv{1-3P_{34}}.
\end{equation}
Substituting $\Psi$ into the projection equation
\begin{equation}
\langle \delta\Psi|(H-E)|\Psi \rangle=0,
\end{equation}
we obtain the coupled integro-differential equation for the
relative function $\chi$ as
\begin{eqnarray}\label{crgm}
\int \left[{\cal H}(\bm R, \bm R')-E{\cal N}(\bm R, \bm
R')\right]\chi(\bm R') d\bm R' =0,
\end{eqnarray}
where the Hamiltonian kernel $\cal H$ and normalization kernel
$\cal N$ can, respectively, be calculated by
\begin{eqnarray}
\left\{
       \begin{array}{c}
          {\cal H}(\bm R, \bm R')\\
          {\cal N}(\bm R, \bm R')
       \end{array}
\right\} =\left<[{\hat \phi}_N(\bm \xi_1,\bm \xi_2 ) {\hat
\phi}_K(\bm \xi_3)]\delta(\bm R-{\bm R}_{NK})\right.\left| \left\{
\begin{array}{c}
          H \\
          1
       \end{array}
\right\}
\right| \nonumber \\
\left.{\cal A}\left[[{\hat \phi}_N(\bm \xi_1,\bm \xi_2) {\hat
\phi}_K(\bm \xi_3)]\delta(\bm R'-{\bm R}_{NK})\right]\right>.
\end{eqnarray}

Eq. $(\ref{crgm})$ is the so-called coupled-channel RGM equation.
Expanding unknown $\chi({\bm R}_{NK})$ by employing well-defined
basis wave functions, such as Gaussian functions, one can solve
the coupled-channel RGM equation for a bound-state problem or a
scattering one to obtain the binding energy or scattering phase
shifts for the two-cluster systems. The details of solving the RGM
equation can be found in Refs.
\cite{kwi77,mka77,mok81,fhuang04nk}.

\section{Results and discussions}

\begin{figure}[htb]
\vglue 2.5cm
 \epsfig{file=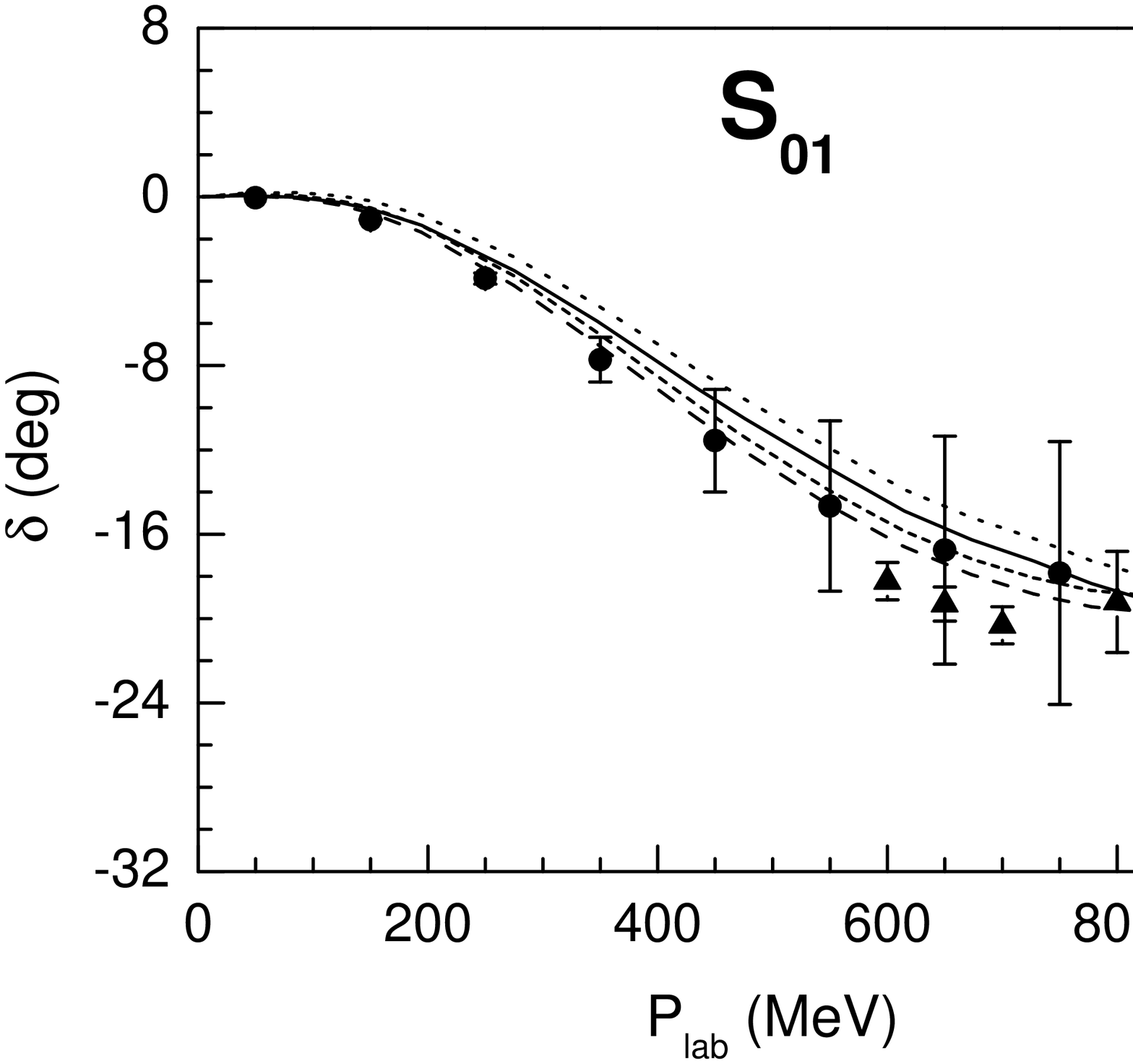,width=7.7cm}
\epsfig{file=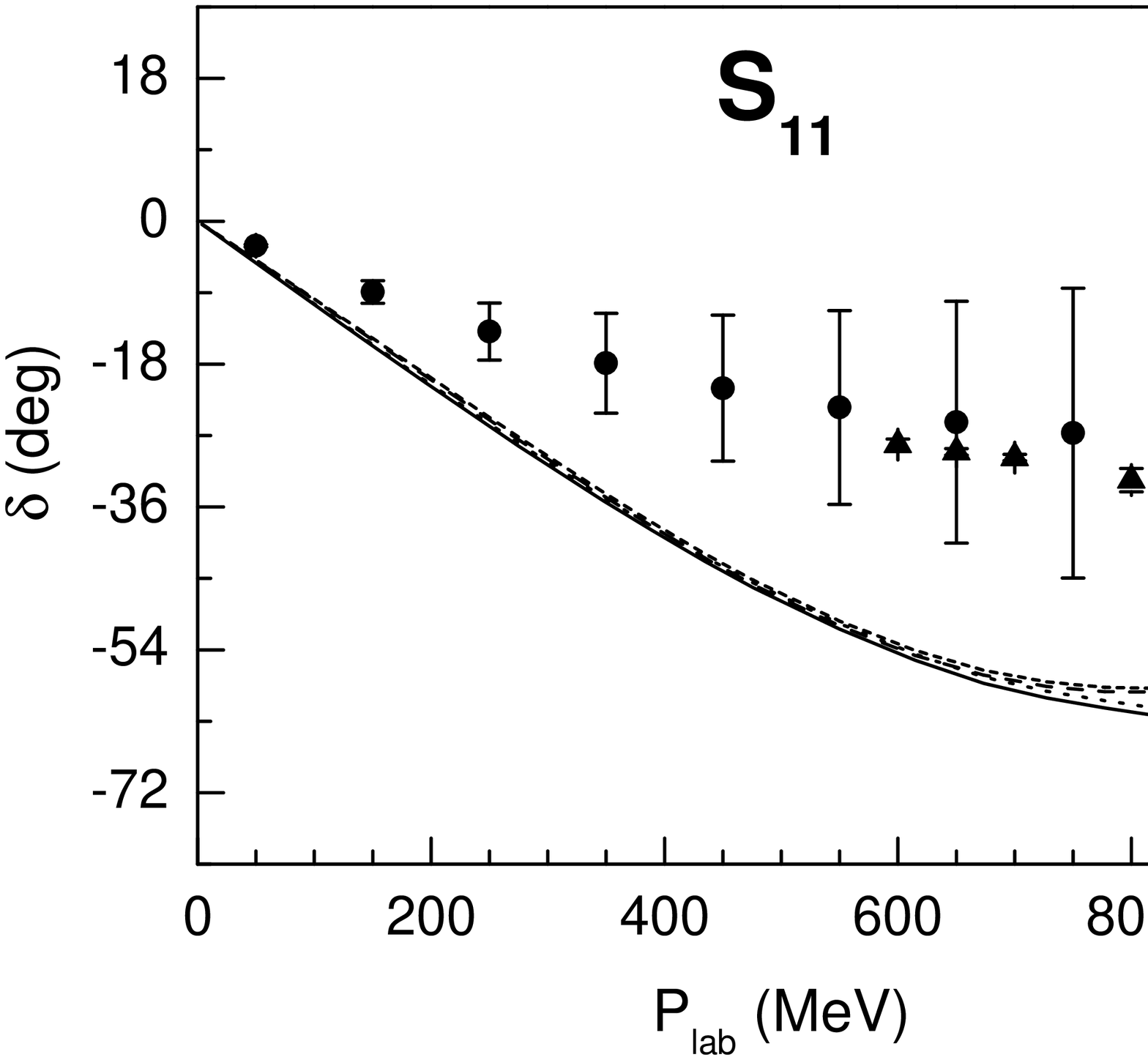,width=7.7cm} \vglue -2.5cm \caption{\small
\label{s0s1} $KN$ $S$-wave phase shifts as a function of the
laboratory momentum of kaon meson. The solid and dotted curves
represent the results obtained in the extended chiral SU(3) quark
model by considering $\theta^S=35.264^\circ$ and $-18^\circ$,
respectively. The dashed and short-dashed curves show the phase
shifts of the original chiral SU(3) quark model by taking
$\theta^S$ as $35.264^\circ$ and $-18^\circ$, respectively.
Experimental phase shifts are taken from Refs. \cite{jsh92}
(circles) and \cite{kha84} (triangles).}
\end{figure}

\begin{figure}[htb]
\vglue 2.5cm \epsfig{file=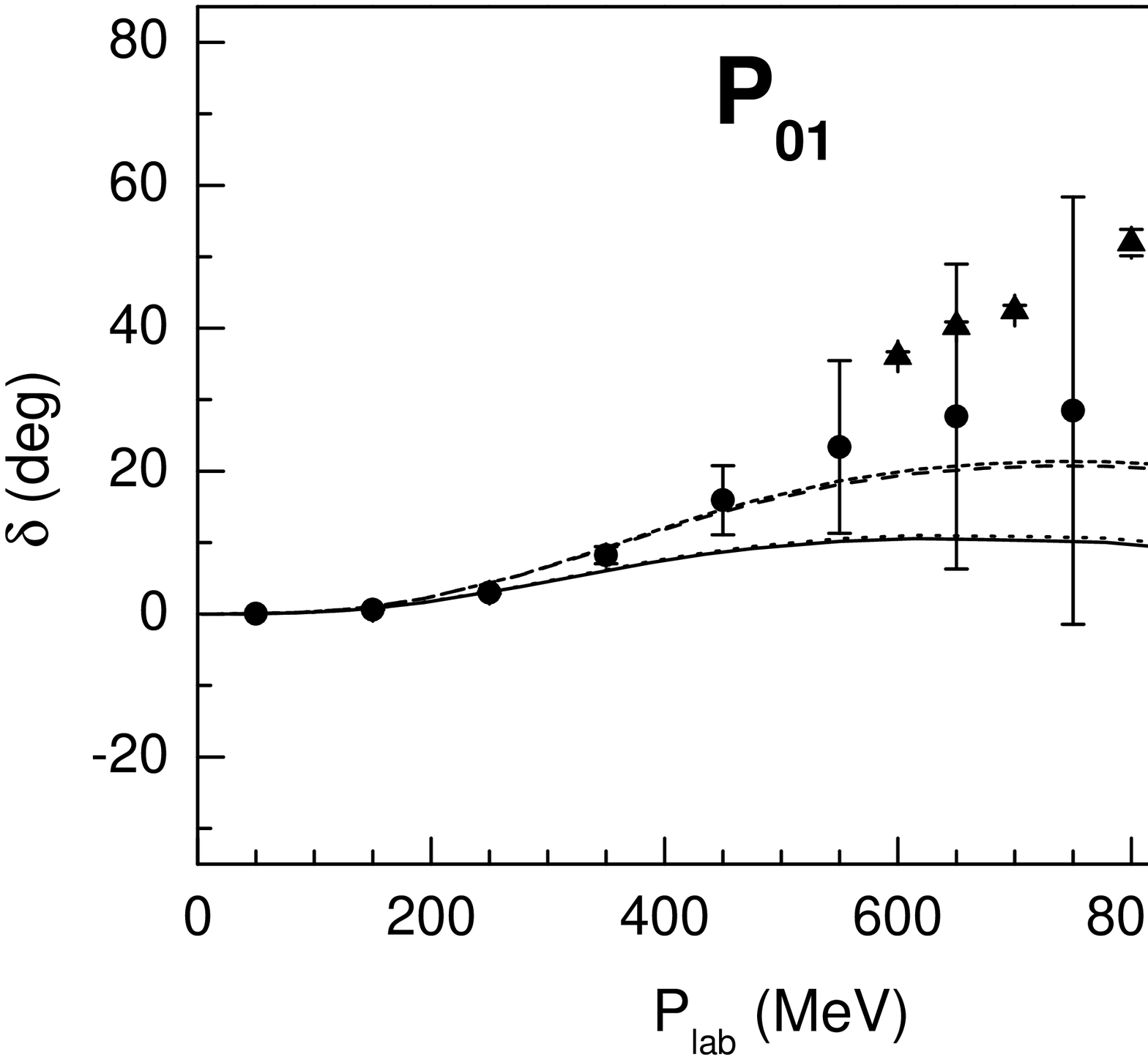,width=7.7cm}
\epsfig{file=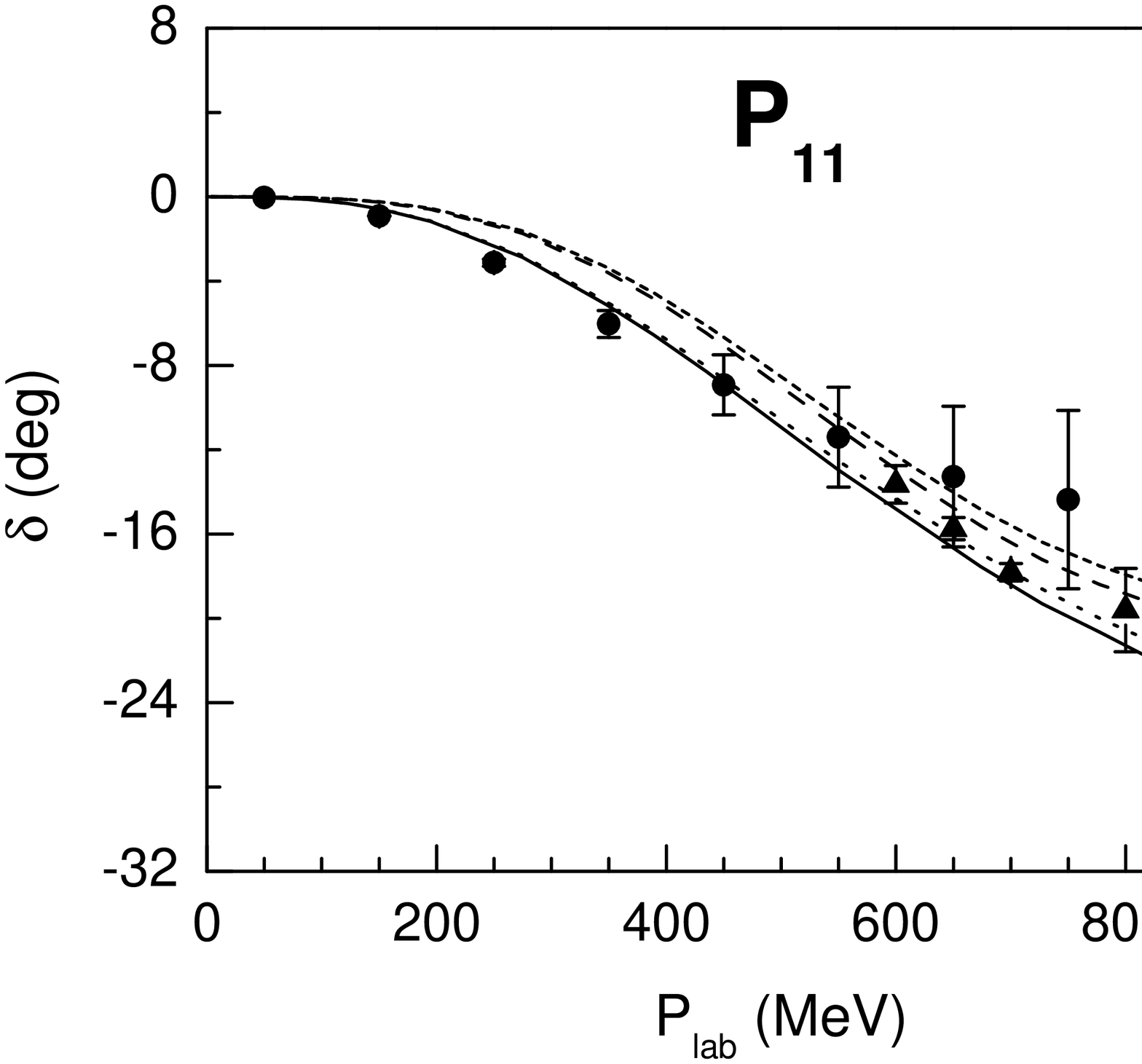,width=7.7cm}
\epsfig{file=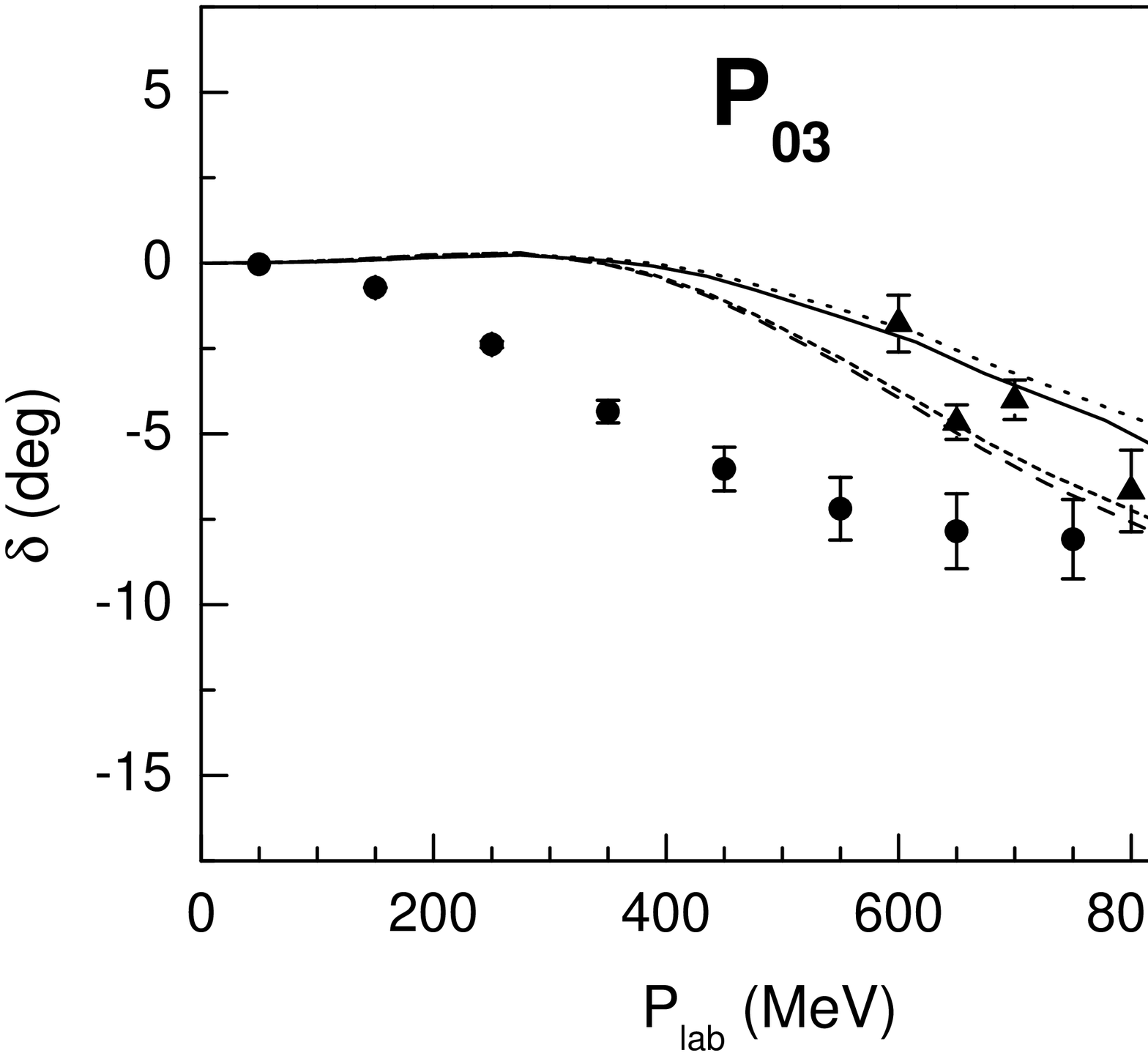,width=7.7cm}
\epsfig{file=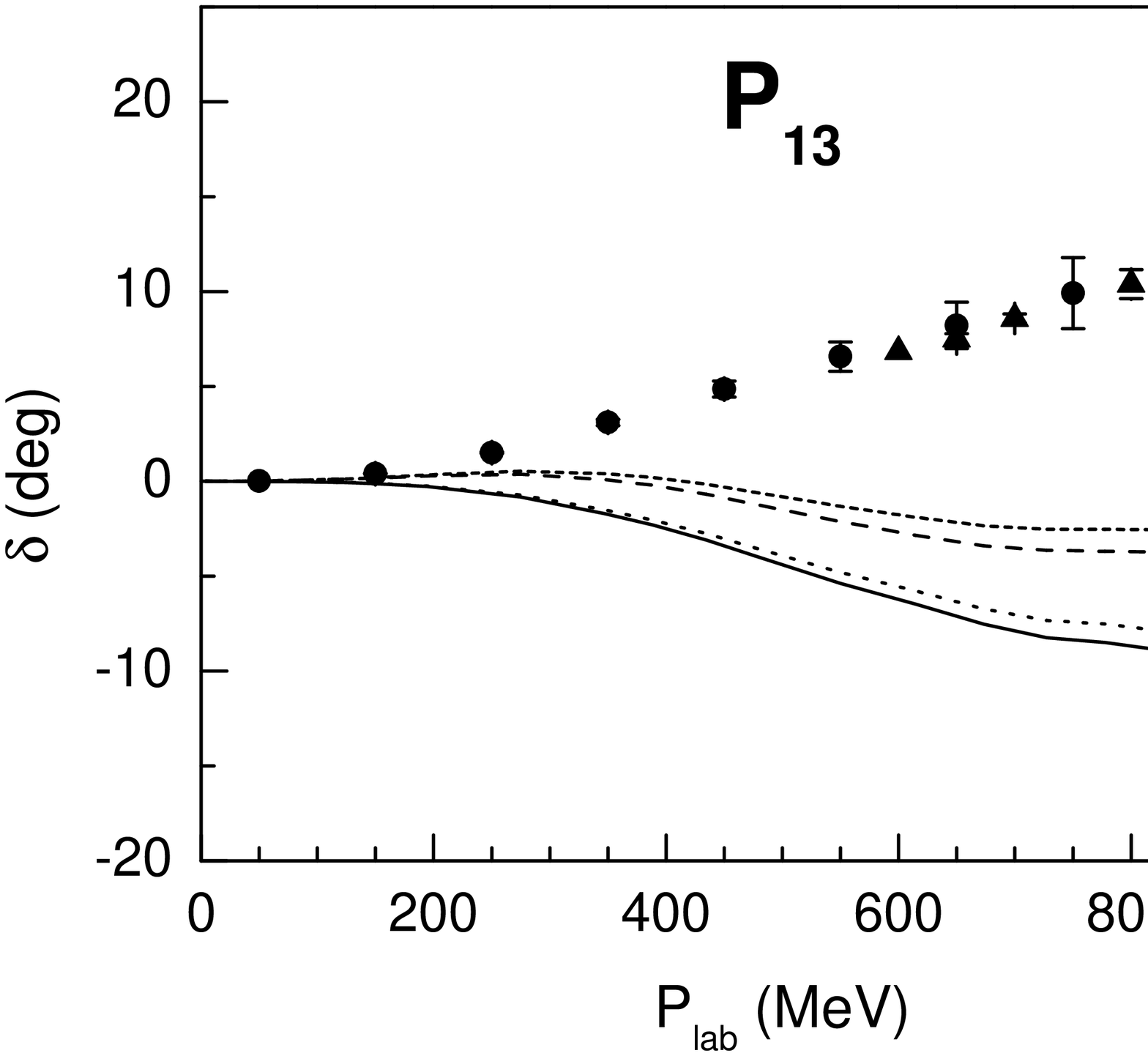,width=7.7cm} \vglue -2.5cm \caption{\small
\label{p0p1} $KN$ $P$-wave phase shifts. Same notation as in Fig.
\ref{s0s1}.}
\end{figure}

\begin{figure}[htb]
\vglue 2.5cm \epsfig{file=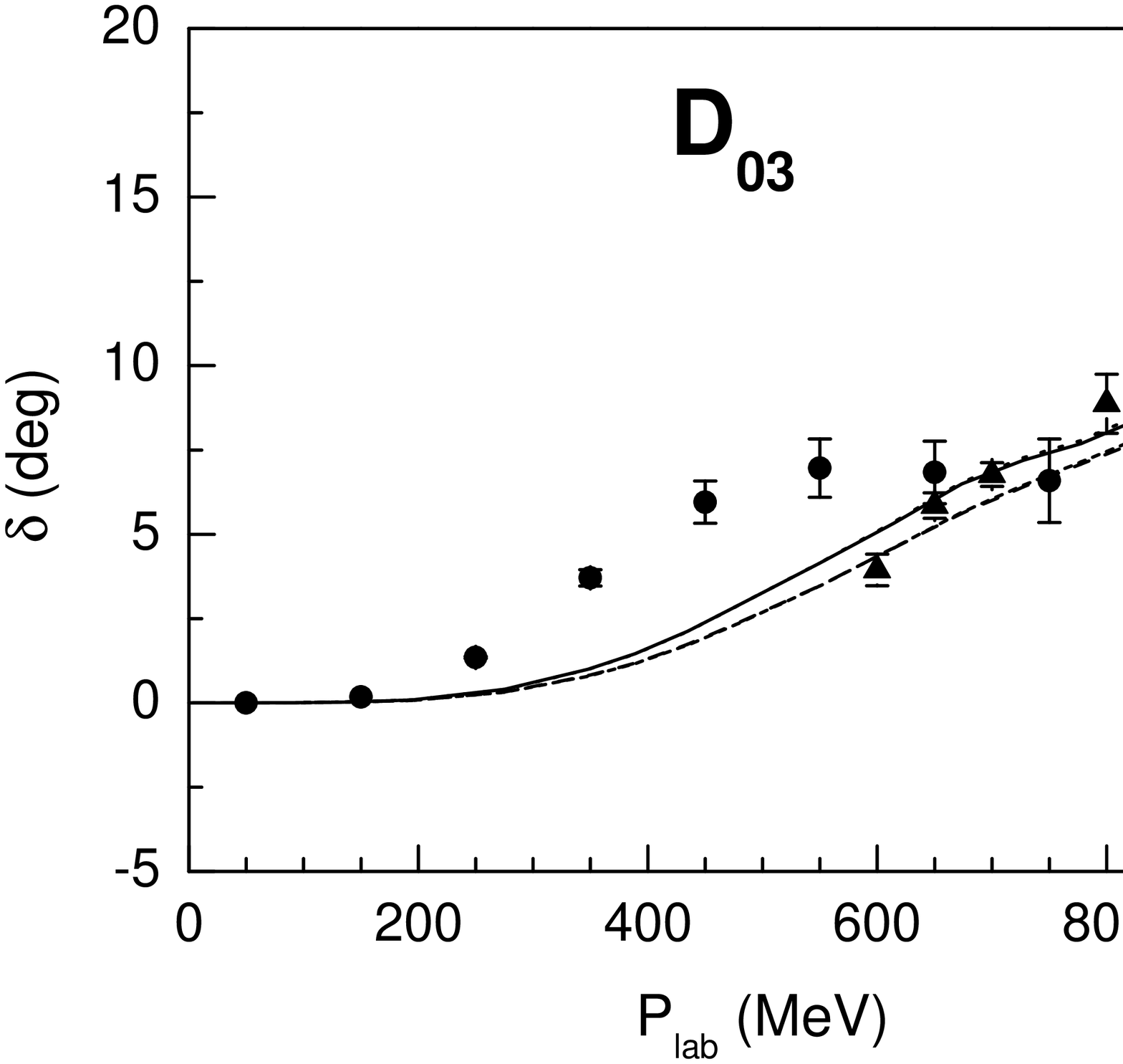,width=7.7cm}
\epsfig{file=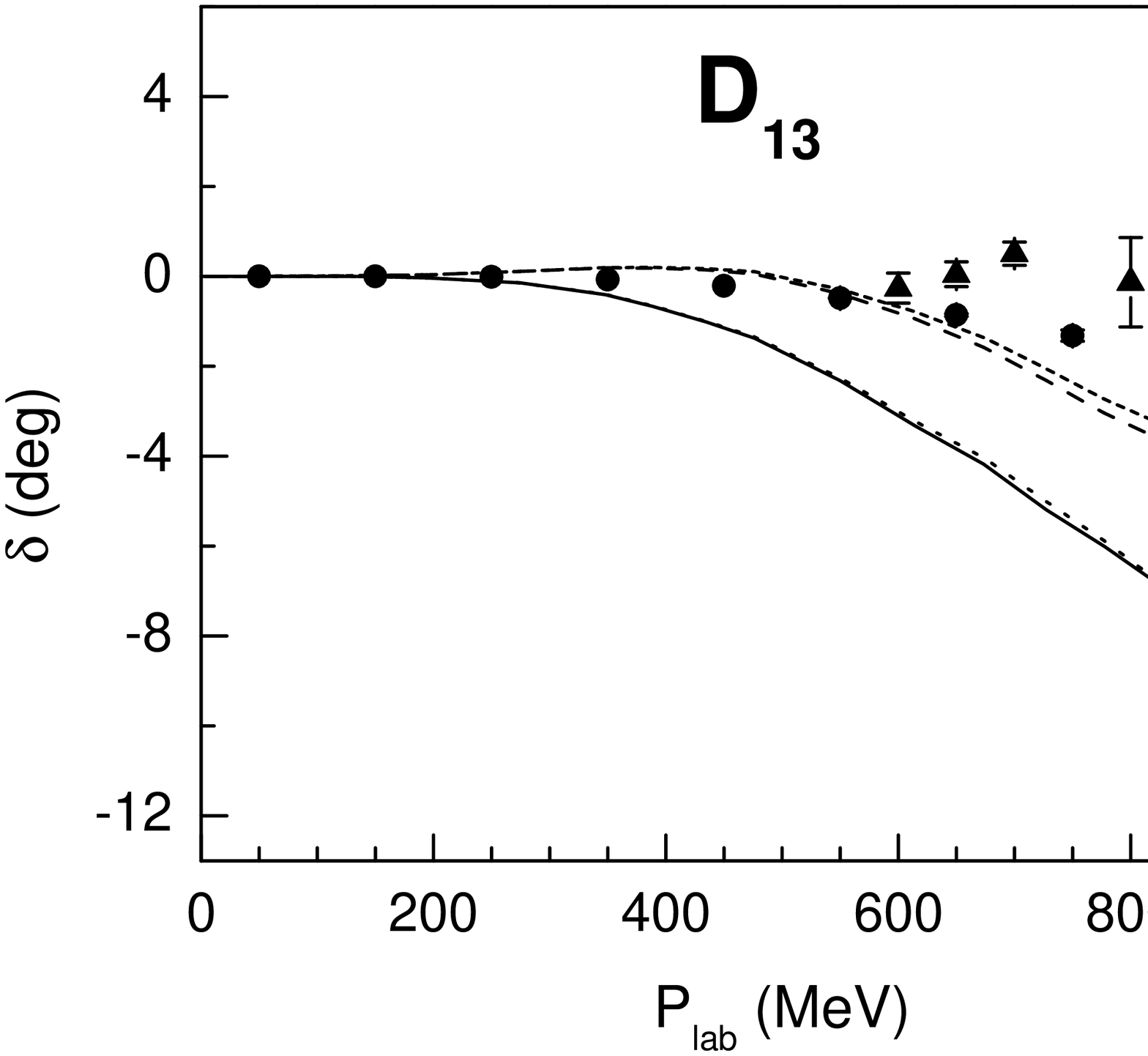,width=7.7cm}
\epsfig{file=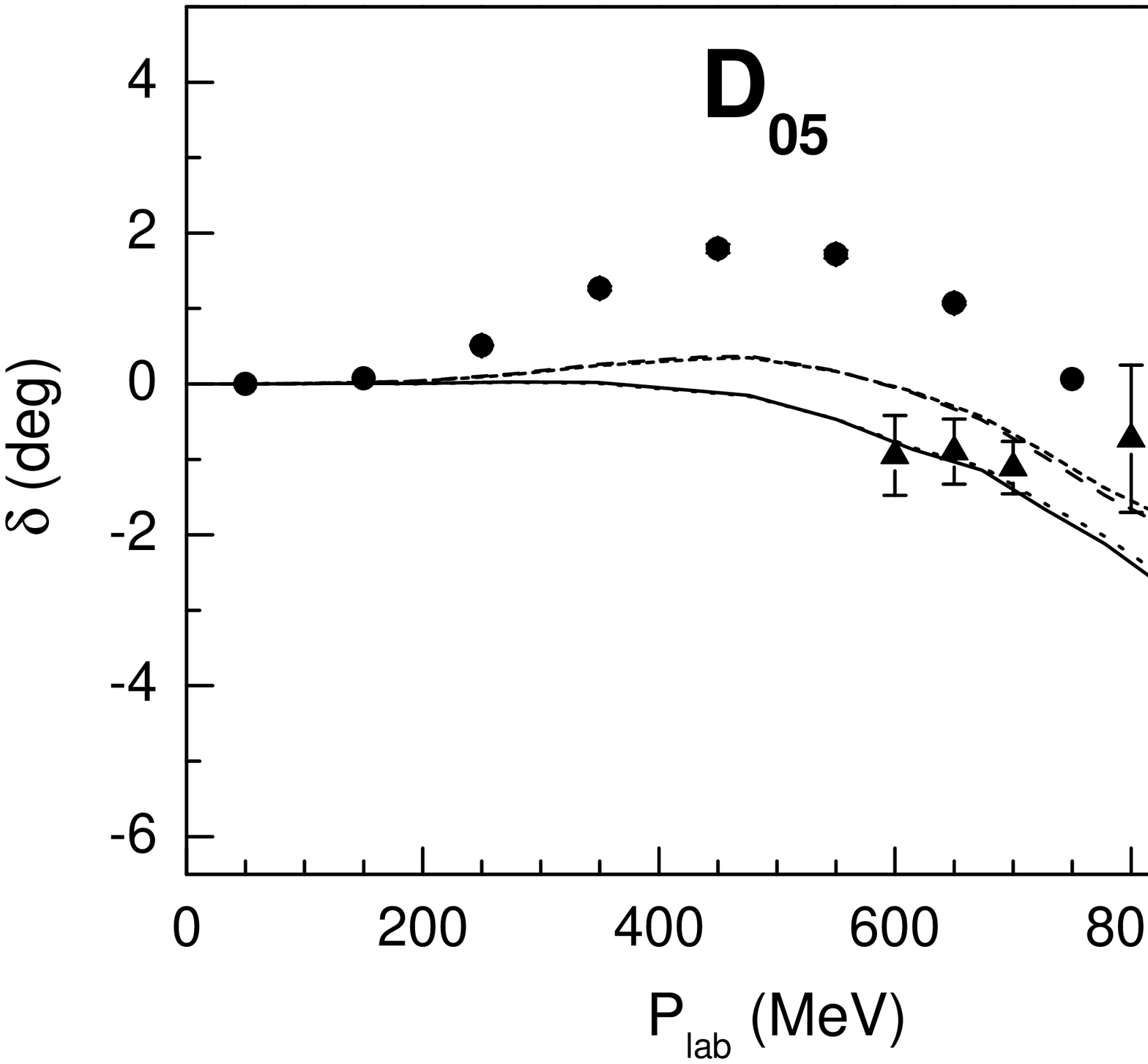,width=7.7cm}
\epsfig{file=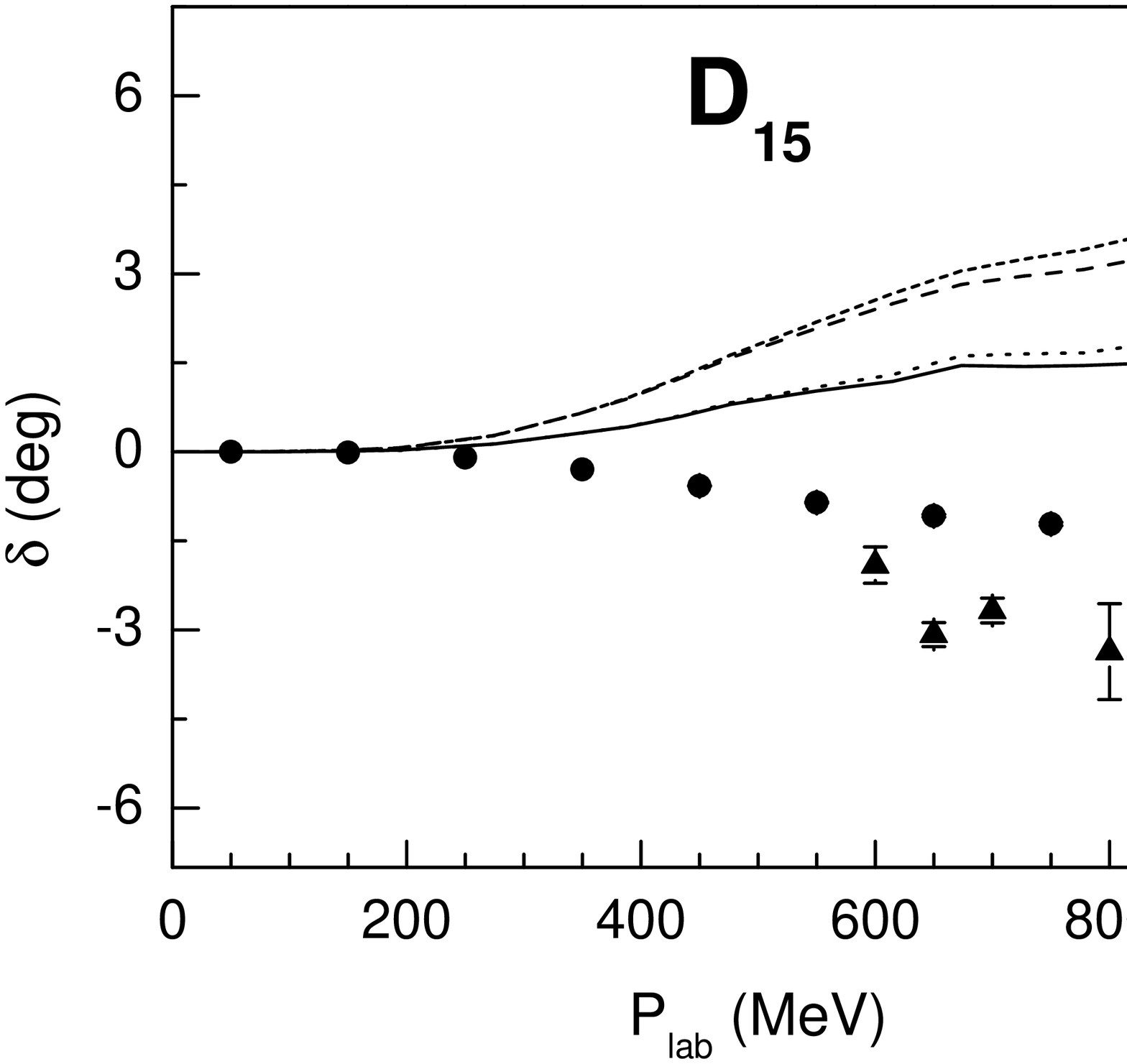,width=7.7cm} \vglue -2.5cm \caption{\small
\label{d0d1} $KN$ $D$-wave phase shifts. Same notation as in Fig.
\ref{s0s1}.}
\end{figure}

\begin{figure}[htb]
\vglue 2.5cm \epsfig{file=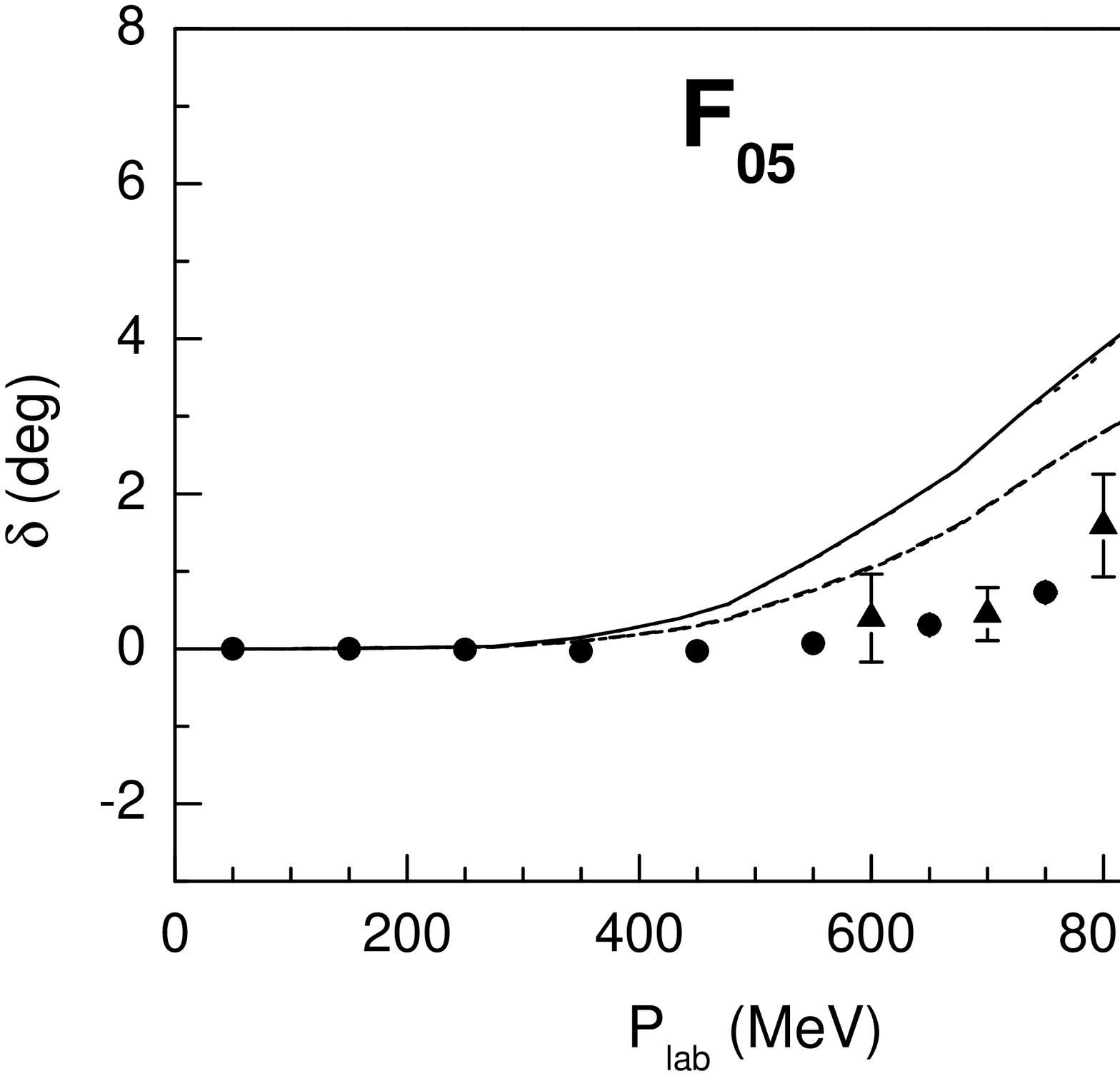,width=7.7cm}
\epsfig{file=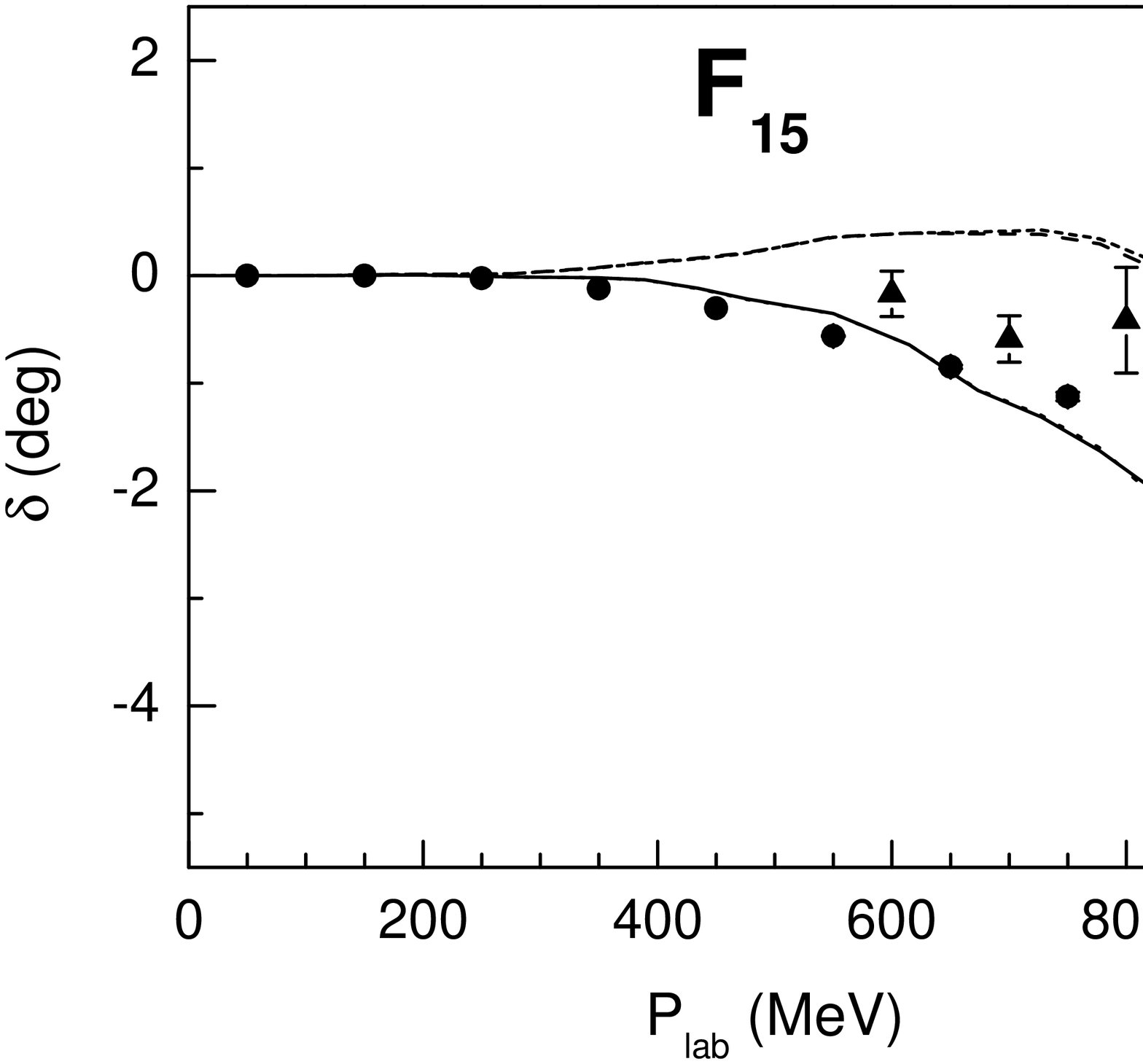,width=7.7cm}
\epsfig{file=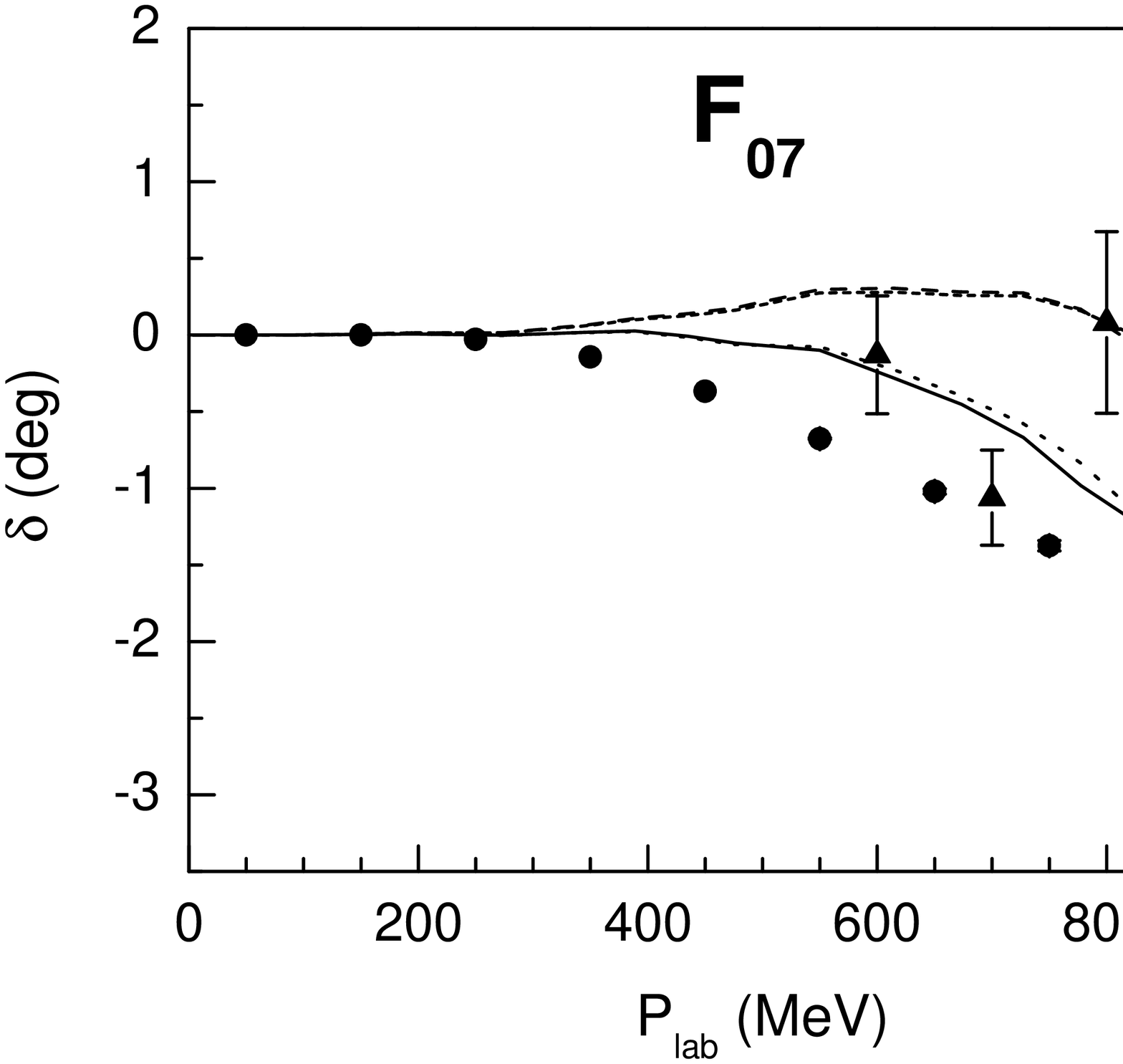,width=7.7cm}
\epsfig{file=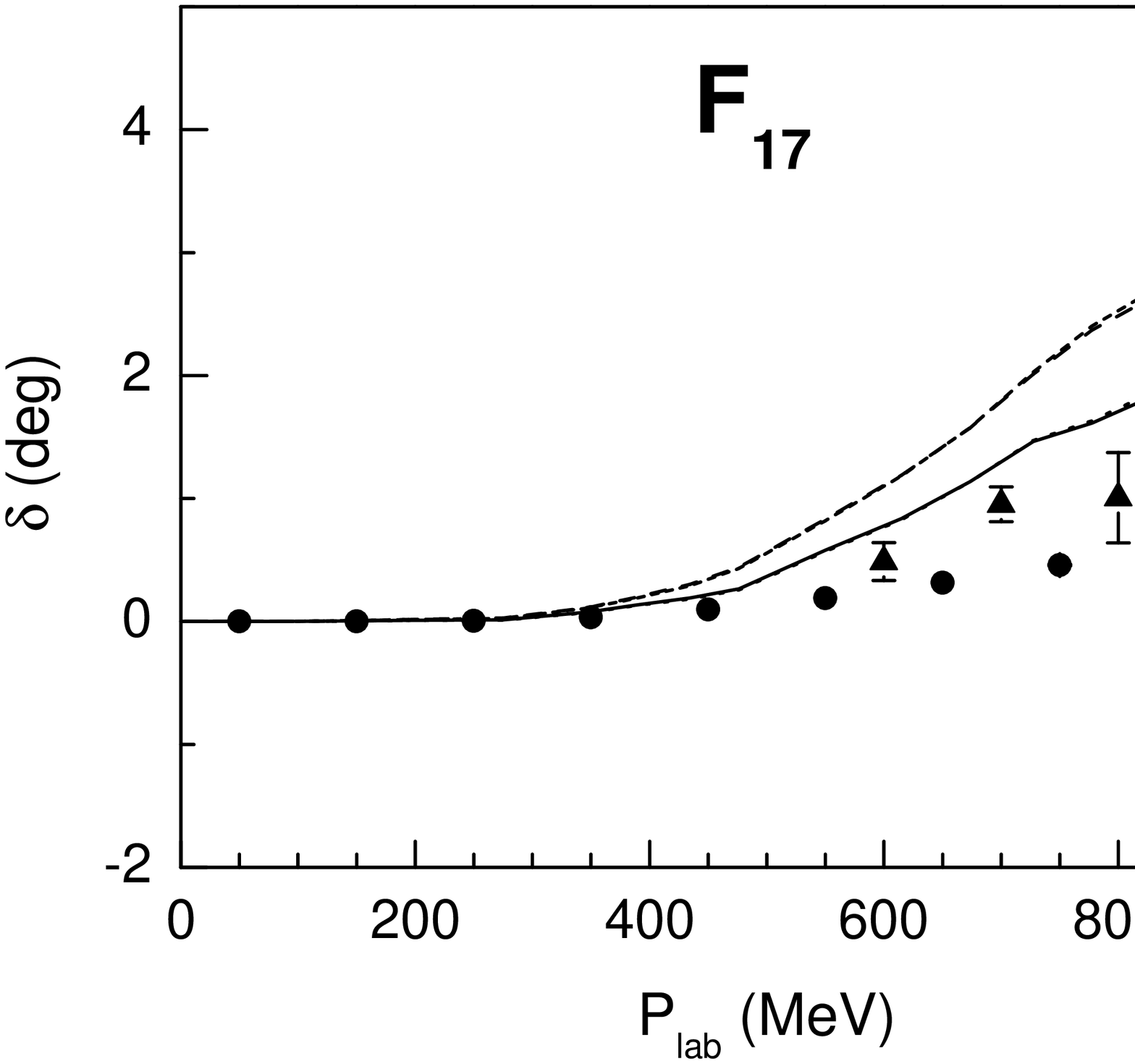,width=7.7cm} \vglue -2.5cm \caption{\small
\label{f0f1} $KN$ $F$-wave phase shifts. Same notation as in Fig.
\ref{s0s1}.}
\end{figure}

\begin{figure}[htb]
\epsfig{file=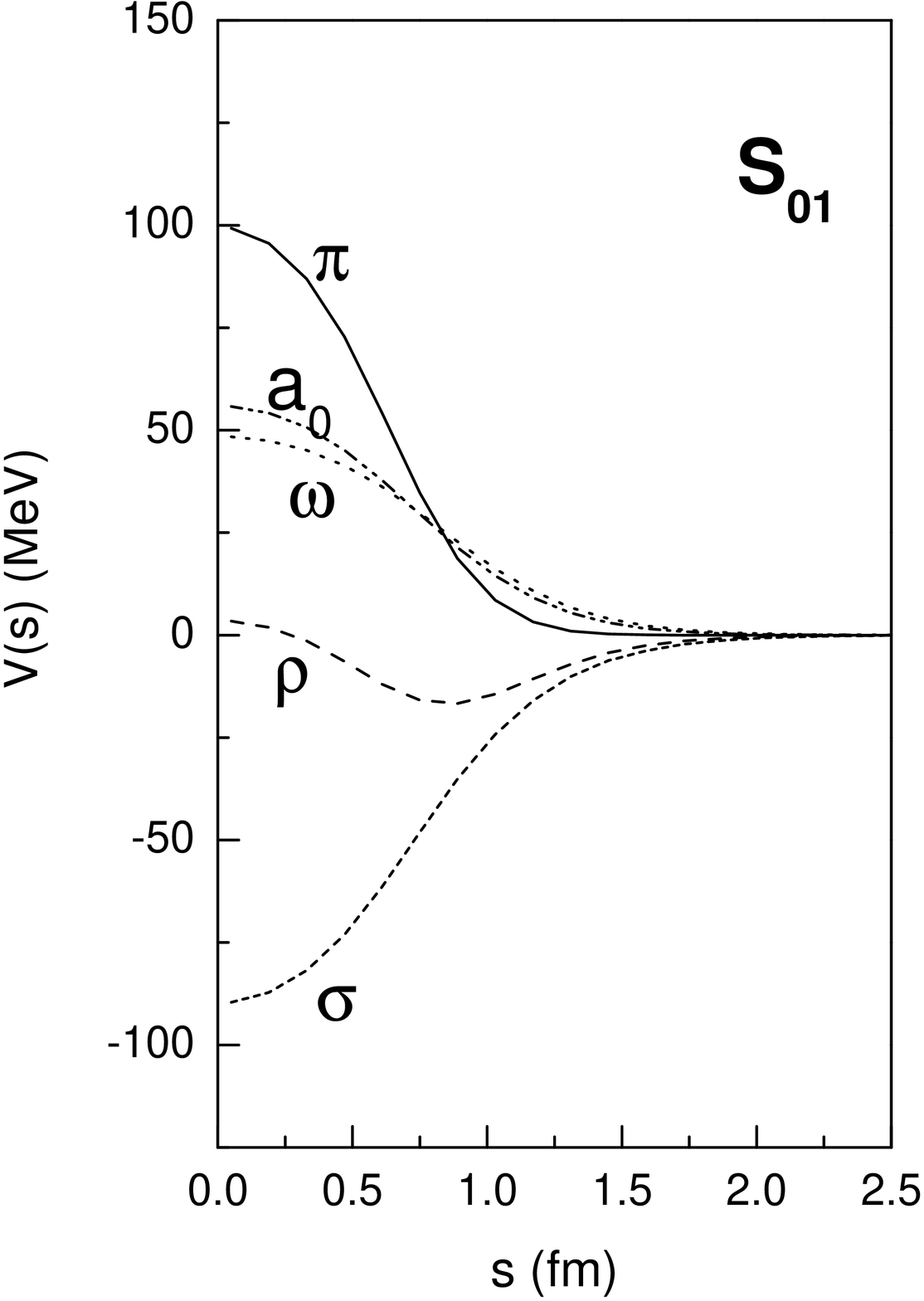,width=7.0cm,height=8.0cm}
\epsfig{file=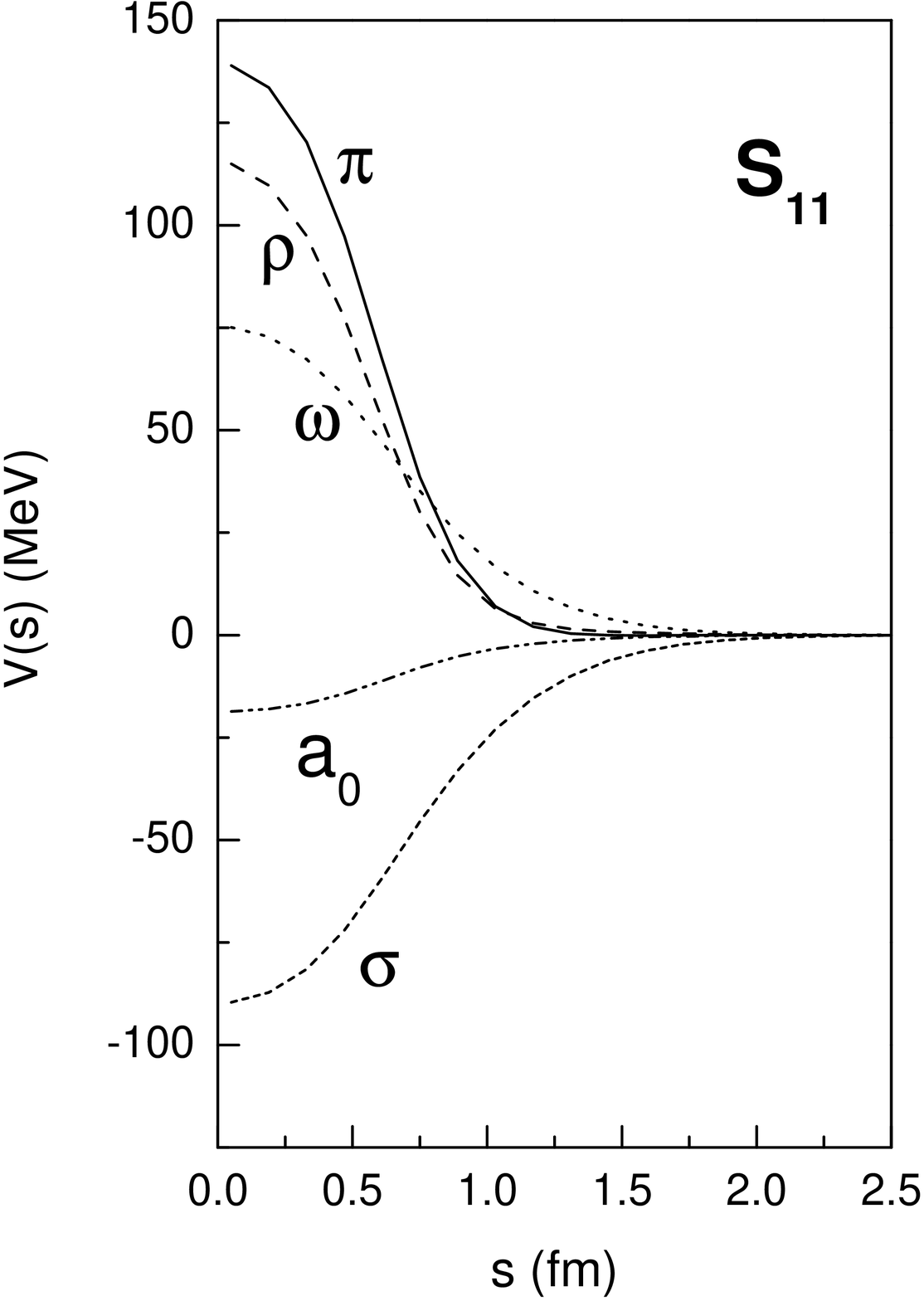,width=7.0cm,height=8.0cm} \vglue
-0.5cm \caption{\small \label{potential} The GCM matrix elements
of $\sigma$, $a_0$, $\pi$, $\rho$, and $\omega$ exchanges in the
extended chiral SU(3) quark model.}
\end{figure}

In the extended chiral SU(3) quark model, the coupling of quarks
and vector-meson field is considered, and thus the coupling
constants of OGE are greatly reduced by fitting the mass
difference between $N$, $\Delta$ and $\Lambda$, $\Sigma$. From
Table \ref{para}, one can see that for both set I and set II,
$g_u^2=0.0748$ and $g_s^2=0.0001$, which are much smaller than the
values of the original chiral SU(3) quark model ($g_u^2=0.7704$
and $g_s^2=0.5525$). This means that the OGE, which plays an
important role of the $KN$ short-range interaction in the original
chiral SU(3) quark model, is now nearly replaced by the
vector-meson exchanges. In other words, in the $KN$ system the
mechanisms of the quark-quark short-range interactions of these
two models are totally different.

A RGM dynamical calculation of the $S$-, $P$-, $D$-, and $F$-wave
$KN$ phase shifts with isospin $I=0$ and $I=1$ is performed, and
the numerical results are shown in Figs. 1--4. Here we use the
conventional partial wave notation: the first subscript denotes
the isospin quantum number and the second one twice of the total
angular momentum of the $KN$ system. For comparison the phase
shifts calculated in the original chiral SU(3) quark model are
also shown in these figures.

Let's first concentrate on the $S$-wave results (Fig. \ref{s0s1}).
In a previous quark model study \cite{bsi97} where the $\pi$ and
$\sigma$ boson exchanges as well as the OGE and confining
potential are taken as the quark-quark interaction, the authors
concluded that a consistent description of the $S$-wave $KN$ phase
shifts in both isospin $I=0$ and $I=1$ channels simultaneously is
not possible. Another recent work in a constituent quark model
based on the RGM calculation gave an opposite sign of the $S_{01}$
channel phase shifts \cite{sle03}. From Fig. \ref{s0s1} one can
see that we obtain a successful description of the $S_{01}$
channel phase shifts, and for the $S_{11}$ partial wave, similar
to that obtained in the original chiral SU(3) model, the trend of
the theoretical phase shifts is also in agreement with the
experiment. Since there is no contribution coming from the
spin-orbit coupling in the $S$-wave, only the central force of the
quark-quark interaction can enter in the scattering process, thus
it plays a dominantly important role. To understand the
contributions of various chiral fields to the $KN$ interaction, in
Fig. \ref{potential} we show the central force diagonal matrix
elements of the generator coordinate method (GCM) calculation
\cite{kwi77} of the $\sigma$, $a_0$, $\pi$, $\rho$, and $\omega$
boson exchanges in the extended chiral SU(3) quark model, which
can describe the interaction between two clusters $N$ and $K$
qualitatively. In Fig. \ref{potential}, $s$ denotes the generator
coordinate and $V(s)$ is the effective boson-exchange potential
between the two clusters. Form this figure we can see that the
$\sigma$ exchange always offers attraction and $\omega$ exchange
offers repulsion in both isospin $I=0$ and $I=1$ channels. This is
reasonable since the $\sigma$ and $\omega$ exchanges are isospin
independent. Contrarily, the $a_0$, $\pi$, and $\rho$ exchanges
are isospin dependent. In the $S_{01}$ partial wave the $a_0$
exchange offers repulsive and $\rho$ exchange offers a little
attractive, while in the $S_{11}$ partial wave the $a_0$ exchange
offers a little attraction and $\rho$ exchange offers repulsion.
In both of these two channels the $\pi$ exchange, existing due to
the quark exchange required by the Pauli principle, always offers
much strong repulsion though the repulsion strength is different.
This means that the one-pion exchange is important and cannot be
neglected on the quark level, which is quite different from the
works on the hadron level where the one-pion exchange is absent in
the $KN$ interaction.

Now look at the $P$-wave $KN$ phase shifts (Fig. \ref{p0p1}). The
results for the $P_{13}$ channel, which are too repulsive in the
original chiral SU(3) quark model when the laboratory momentum of
the kaon meson is greater than 300 MeV, the same case as in
Black's previous work \cite{nbl02}, are now much more repulsive in
the extended chiral SU(3) quark model. For the other channels the
results in both these two models are similar to each other.
Comparing with Ref. \cite{sle03}, we get correct signs and proper
magnitudes of $P_{11}$ and $P_{03}$ waves in both the extended
chiral SU(3) quark model and the original chiral SU(3) quark
model.

For higher-angular-momentum partial waves (Figs.
\ref{d0d1}--\ref{f0f1}), the theoretical phase shifts of $D_{15}$
and $F_{17}$ in the extended chiral SU(3) quark model are improved
in comparison with those obtained from the original chiral SU(3)
quark model, while in the case of the $D_{13}$ the situation is
somewhat less satisfying. For the other channels, the trends of
the calculated phase shifts in both these two models are all in
qualitative agreement with the experiment. Comparing with Ref.
\cite{sle03}, in both these two models, we can get correct signs
of $D_{13}$, $D_{05}$, $F_{15}$, and $F_{07}$ waves, and for
$D_{03}$ and $D_{15}$ channels we also obtain a considerable
improvement on the theoretical phase shifts in the magnitude.

As discussed in Refs. \cite{fhuang04nk,fhuang04dk}, the
annihilation interaction is not clear and its influence on the
phase shifts should be examined. We omit the annihilation part
entirely to see its effect and find that the numerical phase
shifts only have very small changes. This is because in the $KN$
system the annihilations to gluons and vacuum are forbidden and
$u(d)\bar s$ can only annihilate to $K$ and $K^*$ mesons. This
annihilation part originating from the $S$-channel acts in the
very short range, so that it plays a negligible role in the $KN$
scattering process.

The other thing we would like to mention is that our results of
$KN$ phase shifts are independent of the confinement potential in
the present one-channel two-color-singlet-cluster calculation.
Thus the numerical results will almost remain unchanged even the
color quadratic confinement is replaced by the color linear one.

From the above discussion, one sees that though the mechanisms of
the quark-quark short-range interactions are totally different in
the original chiral SU(3) quark model and the extended chiral
SU(3) quark model, the theoretical $KN$ phase shifts of $S$, $P$,
$D$, and $F$ waves in these two models are very similar to each
other. Comparing with others' previous quark model studies, we can
obtain a considerable improvement for many channels. However, in
the present work the $P_{13}$ and $D_{15}$ partial waves have not
yet been satisfactorily described. In this sense, one can say that
the present quark model still has some difficulties to describe
the $KN$ scattering well enough for all of the partial waves. It
should be studied in future work the possibility of that if there
are some physical ingredients missing in our quark model
investigations, as well as the relativistic effects and the
nonelastic channel effects on the $KN$ phase shifts.

By the way, to study the short-range quark-quark interaction more
extensively, or on the other words, to examine whether the OGE or
the vector meson exchange governs the short range interaction
between quarks, besides the $KN$ systems the $\overline{K}N$ is
also an interesting case, since there is a close connection of the
vector-meson exchanges between the $KN$ and $\overline{K}N$
interactions due to $G$-parity transition. Specially, the
repulsive $\omega$ exchange changes sign for $\overline{K}N$,
because of the negative $G$ parity of the $\omega$ meson, and
becomes attractive. However, one should note that the treatment of
the $\overline{K}N$ channel is more complicated than the $KN$
system since it involves $s$-channel gluon and vacuum
contributions. Still the extension of our chiral quark model to
incorporate the gluon and vacuum annihilations in the
$\overline{K}N$ system would be a very interesting new
development. Investigations along this line are planned for the
future.

\section{Summary}

In this paper, we extend the chiral SU(3) quark model to include
the coupling between quarks and vector chiral field. The OGE which
dominantly governed the short-range quark-quark interactions in
the original chiral SU(3) quark model is now nearly replaced by
the vector-meson exchange. Using this model, a dynamical
calculation of the $S$-, $P$-, $D$-, and $F$-wave $KN$ phase
shifts is performed in the isospin $I=0$ and $I=1$ channels by
solving a RGM equation. The calculated phase shifts of different
partial waves are similar to those given by the original chiral
SU(3) quark model. Comparing with Ref. \cite{sle03}, a recent RGM
calculation in a constituent quark model, we can obtain correct
signs of several partial waves and a considerable improvement in
the magnitude for many channels. Nevertheless, in the present work
we do not obtain a satisfactory improvement for the $P_{13}$ and
$D_{15}$ partial waves, of which the theoretical phase shifts are
too much repulsive and attractive respectively when the laboratory
momentum of the kaon meson is greater than 300 MeV. Further the
effects of the coupling to the inelastic channels and hidden color
channels will be considered and the interesting and more
complicated $\overline{K}N$ system will be investigated in future
work.

\begin{acknowledgements}
This work was supported in part by the National Natural Science
Foundation of China, Grant No. 10475087.
\end{acknowledgements}

\end{document}